\documentclass[sigconf, 10pt, letter]{acmart}

\renewcommand\footnotetextcopyrightpermission[1]{} 
\usepackage{comment}
\usepackage{courier}
\usepackage{algorithmic}
\usepackage{graphicx}
\usepackage{xcolor}
\usepackage{enumitem}
\usepackage{hyperref}
\usepackage{textcomp}
\usepackage[T1]{fontenc}
\usepackage{textcomp}
\usepackage{multirow}
\usepackage{listings}
\usepackage{tikz,url}
\usetikzlibrary{arrows.meta}
\usetikzlibrary{positioning}
\usetikzlibrary{shapes}

\usepackage{amsmath}
\usepackage{fancyhdr}
\usepackage[linguistics]{forest}

\definecolor{codegreen}{rgb}{0,0.6,0}
\definecolor{codegray}{rgb}{0.5,0.5,0.5}
\definecolor{codepurple}{rgb}{0.58,0,0.82}
\definecolor{backcolour}{rgb}{0.95,0.95,0.92}

\lstdefinestyle{mystyle}{ 
    backgroundcolor=\color{backcolour},   
    commentstyle=\color{codegreen},
    keywordstyle=\color{magenta},
    numberstyle=\tiny\color{codegray},
    stringstyle=\color{codepurple},
    basicstyle=\ttfamily\footnotesize,
    breakatwhitespace=false,          
    breaklines=true,                 
    captionpos=b,                    
    keepspaces=true,                 
    numbers=left,                    
    numbersep=5pt,                  
    showspaces=false,                
    showstringspaces=false,
    showtabs=false,                  
    tabsize=2
}
\begin{document}

\title{Exploitation and Sanitization of Hidden Data in PDF Files}
\subtitle{Do Security Agencies Sanitize Their PDF files?}

\author{Supriya Adhatarao}

\affiliation{%
  \institution{Univ. Grenoble Alpes, Inria- France}
}
\email{supriya.adhatarao@inria.fr}

\author{C\'edric Lauradoux}
\affiliation{%
\institution{Univ. Grenoble Alpes, Inria- France}
%  \institution{INRIA Grenoble Rhône-Alpes}
%  \city{Montbonnot Saint-Martin}
  %\country{France}
  }
\email{cedric.lauradoux@inria.fr}

\begin{abstract}
Organizations publish and share more and more electronic documents like PDF files. Unfortunately, most organizations are unaware that these documents can compromise sensitive information like authors names, details on the information system and architecture. All these information can be exploited easily by attackers to footprint and later attack an organization. In this paper, we analyze hidden data found in the PDF files published by an organization. We gathered a corpus of 39664 PDF files published by 75 security agencies from 47 countries. We have been able to measure the quality and quantity of information exposed in these PDF files. It can be effectively used to find weak links in an organization: employees who are running outdated software. We have also measured the adoption of PDF files sanitization by security agencies. We identified only 7 security agencies which sanitize few of their PDF files before publishing. Unfortunately, we were still able to find sensitive information within 65\% of these sanitized PDF files. Some agencies are using weak sanitization techniques: it requires to remove all the hidden sensitive information from the file and not just to remove the data at the surface. Security agencies need to change their sanitization methods. 
\end{abstract}

\keywords{Footprinting, sanitization, hidden data, metadata, PDF files.}

\maketitle
\section{Introduction}

%The Portable Document Format (PDF) is a very popular file format used online. 

Organization footprinting~\cite{Faircloth2017} regroups all the techniques used by hackers to collect as much information as possible about their victims. The goal of this reconnaissance is not only to obtain details on the people but also on the infrastructures they use (network, hardware, system\ldots). Footprinting includes techniques like OS fingerprinting~\cite{Shamsi2014} or the exploitation of all the documents published by the organization~\cite{Alonso2009a,Alonso2009b} using tools like FOCA\footnote{\url{https://github.com/ElevenPaths/FOCA}}. This later technique is particularly attractive for hackers because it is rather inexpensive and effortless. Organizations publish on their websites many Microsoft Office (\texttt{doc}, \texttt{docx}\ldots) or Portable Document Format (PDF) files. All these file formats are particularly interesting for a hacker because they include hidden data which describe the authoring process. The impact of hidden data was highlighted during two events that occurred during the Iraq War.

In February 2003, the British government of Tony Blair published on its website a dossier on Iraq's security and intelligence organizations. The  dossier was a Microsoft Word file\footnote{Retrieved 02/24/2021 at  \url{http://web.archive.org/web/20040329171413/http://www.computerbytesman.com/privacy/blair.doc}}. The file was analyzed by the IT researcher Richard M. Smith\footnote{Retrieved 02/24/2021 on \url{http://web.archive.org/web/20040113074742/http://www.computerbytesman.com/privacy/blair.htm}} who retrieved the revision logs. It was easy to identify the authors and their positions in government from these revision logs. The British government was greatly embarrassed by the information exposed by those hidden information.

In 2005, an incident occurred between American soldiers and Italian Secret Service officers near Baghdad International Airport causing the death of an Italian officer. Multi-National Force-Iraq issued a report  on its investigation of the shooting. That report was posted as an unclassified PDF file with classified sensitive data obscured from public view. However, it was discovered that copying and pasting the classified sections revealed the blocked text.

Hidden data in electronic documents matters and they must not be underestimated. Promptly after these two events, the National Security Agency (NSA) acknowledged in~\cite{NSA2005} that hidden information in documents are a real security threat:
\begin{quote}
\textbf{Meta-data and Document Properties} - In addition to the visible content of a document, most office tools, such as MS Word, contain substantial hidden information about the document.  This information is often as sensitive as the original document, and its presence in downgraded or sanitized documents has historically led to compromise.
\end{quote} 

Sanitization is the obvious choice to deal with hidden data before sharing or publishing a document. However, most organizations are unaware that they need to sanitize their files. In this paper, we investigate how the hidden data of PDF files can be exploited and if they are sanitized. We focus on answering two questions:
\begin{itemize}
\item What can be done using the hidden data of PDF files? 
\item Are there any organizations that sanitize their PDF files?
\end{itemize}

To answer these questions, we have crawled the websites of 75 security agencies of 47 countries and  collected 39664 PDF files. For the majority of the files (76\%), we were able to recover the authoring process: we identify the PDF producer tool and the Operating System (OS) used by the file's authors. Collecting and analyzing PDF files from the same source over several years can reveal the habits of a given employee. It is possible to learn if he/she update/change (or not) their software regularly. For instance, we found one employee of a security agency who has never changed or updated his/her software during a period of 5 years. This kind of information is particularly interesting for a hacker to target an individual with bad software habits. By analyzing the PDF files published by several employees of the same agency, it is possible to learn the software policies of an agency.   We found at least 19 security agencies in our dataset who are using the same software over a period of 2 years or more. Around 38 security agencies have better practices and are regularly changing or updating their software. 

We have also observed that, 24\% of all the PDF files have been sanitized. We have identified 7 agencies which sanitize their PDF files before publishing. However, our analysis shows that the sanitization method used was weak for 65\% of the sanitized PDF files. It was possible to recover sensitive information from these files. Only 3 agencies are reaching a satisfying sanitization level.

This paper is organized as follows. In Section~\ref{sec:hidden}, we present the hidden data included in a PDF file and some related work. The principles of PDF sanitization are given in Section~\ref{sec:sani}. Our results on the sanitization and on hidden data of PDF files  are described respectively in Section~\ref{sec:exploit} and in Section~\ref{sec:dataset}. 
\section{Hidden data in PDF files}\label{sec:hidden}
Adobe Portable Document Format is one of the most widespread and used file format throughout the world. It has been created by Adobe Systems in 1993 to extend Postscript and it is now  standardized as an open format ISO since 2008 and the latest version of the standard is PDF 2.0~\cite{Iso32000}. 
Even though this format has been evolved over time and later versions have been improved  to provide better security, it is still not immune to flaws and attacks that could be misused. 
Many works have been done in the past on PDF file security~\cite{Stevens2011,Smutz2012,Carmony2016,Xu2016,Maiorca2019} and privacy~\cite{Smith2000,Martin2003,Garfinkel2014}. Some works also include results on PDF file sanitization~\cite{Aura2006,Garfinkel2014,Feng2018}.

The visible content of a PDF file can directly reveal many information on the authors: their names, their organization\ldots It can embed many different media: text, pictures, videos\ldots During our analysis of PDF files, we observed that a PDF file is basically collection of indirect objects. There are eight types of objects: boolean, numbers, strings, names, arrays, dictionaries, streams and the null object. Using these objects, data is stored within the PDF file. Sometimes objects in a PDF file can include hidden data that is not displayed while viewing the PDF but can be extracted using appropriate tools. 

NSA provides a list of  eleven main types of hidden data and embedded content that may be found in PDF files~\cite{NSA2008,NSA2011}. Only after all these eleven types of hidden information is removed, a PDF file is safe for distribution.

\begin{enumerate}
\setlength\itemsep{0.4em}
\item Metadata
\item Embedded Content and Attached Files
\item Scripts
\item Hidden Layers
\item Embedded Search Index
\item Stored Interactive Form Data
\item Reviewing and Commenting
\item Hidden Page, Image, and Update Data
\item Obscured Text and Images
\item PDF Comments (Non-Displayed)
\item Unreferenced Data
\end{enumerate}
 
We explain the most commonly found types of hidden data in the rest of this section.
\subsection{Dynamic Resources}

It was demonstrated in~\cite{Castiglione2010} that a PDF can be tracked when it is read. A malicious author can use this technique to obtain information on his/her reviewer during the review process of a conference. It is based on the possibility to embed scripts that are executed by the PDF viewer when the file is displayed. The script needs to download external resources controlled by the author. It exposes the IP address, user-agent and the time of the request of the reviewer. It was even discovered in 2019 that embedding a script was not needed to track a PDF file. CVE-2019-8097 (see \url{https://nvd.nist.gov/vuln/detail/CVE-2019-8097}) shows that it is possible to insert an URI in a PDF file which is  automatically downloaded by certain PDF viewers. This is the application of web bugs~\cite{Smith2000,Martin2003,Vaidya2017} applied to PDF files. 

The dangers of scripts and dynamic resources in PDF files are  well-known and many solutions exist (see~\cite{Laskov2011,Stevens2011,Srndic2013,Maiorca2019a,Maiorca2019b})  and the method proposed in~\cite{Castiglione2010} is unlikely to work anymore. The exploitation of dynamic resources requires direct interaction with the targeted organization, \emph{(active footprinting)}. It is out-of-scope of this paper which focuses only on the \emph{passive footprinting}.

\subsection{Paths and Images}
Images embedded in a PDF file can provide many information on the authors. If an author cleans the metadata of the PDF document but forgets to clean the metadata of the embedded image files, one can still extract information on the author. Image metadata can include author name, username, path from where the file was inserted\ldots Image metadata can include information on creation and modification of the image which can reveal different software used and the habits of the author. Geo-location features stores the geo-data revealing the location where the image was taken which may impacts the security and privacy of an individual.

Several tools like \texttt{ImageMagick}\footnote{\url{https://imagemagick.org/script/download.php}}, \texttt{exiftool} and \texttt{exiv2}\footnote{\url{http://manpages.ubuntu.com/manpages/hirsute/en/man1/exiv2.1.html}} can be used to view metadata. We have used \texttt{pdfxplr}\footnote{https://github.com/sowdust/pdfxplr} in our work to extract the \texttt{Username} and \texttt{PATH} information from the image files embedded within the PDF documents. Listing~\ref{path} shows the \texttt{PATH} extracted from one of the PDF files containing a image file. We can see that the author is providing the location where the image file was stored along with the username.

\begin{center}
\lstset{ %
    basicstyle=\ttfamily\footnotesize,
    frame=single,
    keywordstyle=\color{blue},
    language=Bash,
    showstringspaces=false,
    caption={A PDF object displaying the PATH information of an embedded image file.},
    captionpos=b,
    label=path,
    morekeywords={blue},
}
\begin{minipage}{3.31in}
\begin{lstlisting}[basicstyle=\small]
19693 0 obj
<<
/K 29/P 19690 0 R/S/InlineShape/Alt(Description: C:\\Users\\Mazhar\\Desktop\\scml.JPG)/Pg 19761 0 R
>>
endobj
\end{lstlisting}
\end{minipage}
\end{center}

Image extraction and analysis was proven to be successful in~\cite{Aura2006,Feng2018} to recover information on the authors of the PDF file. Aura \textit{et al.}~\cite{Aura2006} have tested 43 anonymous PDF submissions of a conference to detect if any leakage was present. They found 3 submissions which were not properly anonymized. One submission contained a metadata field with the authors names and the two others had image files with identifying metadata. They proposed a tool based on regular expression to detect 
usernames, device or organization names, identifiers\ldots in a PDF file. 

Feng \textit{et al.}~\cite{Feng2018} have also proposed a tool to detect privacy leakages in PDF files with a focus on the extracted images. The main difference between their tool and the tool of Aura \textit{et al.}~\cite{Aura2006} is the use of  text  mining and information  retrieval  and  natural language processing to detect identifying information. %We have not considered this problem in this paper.

\subsection{Comments, annotations and other objects}
Comments can be inserted anywhere in the PDF file by starting a line with \texttt{\%} symbol, PDF viewers just ignore the comments and display the content stored in PDF objects. Comments are often a good source of information. 

PDF files sometimes include annotations, these annotations can stay invisible when a PDF file is viewed. However,  they can be retrieved using simple string extraction commands like \texttt{strings}. Often authors forget to remove the annotations from their files and it can expose their identity. Listing~\ref{annot} shows an object extracted from a PDF file that reveals the annotated message: \texttt{changes made by john doe} and also the Username: \texttt{sab}.  

\begin{center}
\lstset{ %
    basicstyle=\ttfamily\footnotesize,
    frame=single,
    keywordstyle=\color{blue},
    language=Bash,
    showstringspaces=false,
    caption={Annotation object of a PDF file containing the message and username.},
    captionpos=b,
    label=annot,
    morekeywords={blue},
}
\begin{lstlisting}[basicstyle=\small]
152 1 obj 
<<
/Type /Annot /Rect [165.897704918 615.5800226116 189.897704918 639.5800226116 ] /Subtype /Text /M (D:20210225232546) /C [1 1 0 ] /Popup 153 1 R /T (\FE\FF\00s\00a\00b) /P 3 0 R /Contents (\FE\FF\00c\00h\00a\00n\00g\00e\00s\00 \00m\00a\00d\00e\00 \00b\00y\00 \00j\00o\00h\00n\00 \00d\00o\00e) 
>> 
endobj
\end{lstlisting}
\end{center}

Apart from comments and annotations, some specific objects within a PDF file can also include some sensitive information.  We found that registry objects, font objects\ldots can include metadata information. Listing~\ref{t3} shows two objects extracted from PDF files that reveal information about the OS and other software used.

\begin{center}
\lstset{ %
    basicstyle=\ttfamily\footnotesize,
    frame=single,
    keywordstyle=\color{blue},
    language=Bash,
    showstringspaces=false,
    caption={Specific objects revealing information on the authoring process of a PDF file.},
    captionpos=b,
    label=t3,
    morekeywords={blue},
}
\begin{minipage}{3.5in}
\begin{lstlisting}[basicstyle=\small, ]
15 0 obj
<<
/DW 1000/CIDSystemInfo
<<
/Supplement 0/Registry(Adobe)/Ordering(Identity)
>>
/Subtype/CIDFontType2/BaseFont/CAFBBG+TimesNewRomanPSMT/Type/Font/FontDescriptor 23 0 R/W[267[610 443] 284[333]]
>>
endobj


1459 0 obj
<<
/Platform(Macintosh)/Creator(FileMaker Pro Advanced 14.0.1)/DLI_Copyright(Datalogics Interface \(DLI\) Copyright \(C\) 1998-2012 Datalogics, Inc. -- www.datalogics.com)/Producer(Adobe PDF Library 10.1; modified using iText 2.1.7 by 1T3XT)/Title()/Keywords()/ModDate(D:20180223153614+01'00')/Subject()/DLI(10.1.0.50)/Author()/CreationDate(D:20180220152519+01'00')
>>
endobj
\end{lstlisting}
\end{minipage}
\end{center}

% (but if used carefully, it's almost undetectable). Also, in \cite{corkami} Ange A. achieved to create a \code{PDF} image when applied a \code{3DES} encryption algorithm obtains a valid \code{JPG} file.

%Past works have also shown that it is possible to hide information using these features. In~\cite{distributed_steganography_in_multiple_pdf}, authors

\subsection{Metadata}
The metadata are often used in cataloging to help searching for documents in external databases. Metadata are stored either in a document information dictionary or in a metadata stream. When the metadata are stored in a document information dictionary, the PDF file's trailer include an optional \texttt{Info} entry that holds the document information dictionary consisting of metadata information associated to the PDF file. Metadata information can  also be stored in streams called metadata stream. It is  either for the complete PDF file or for some components within the PDF file. Metadata stream is represented using Extensible Markup Language (XML).

Many studies in the past have discussed on the impact of PDF metadata~\cite{Alonso2009a,Alonso2009b,Smutz2012,Garfinkel2014,Mendelman2018}. 
There are several ways and tools that could be used to view and extract metadata information from PDF files: \texttt{exiftool}\footnote{https://exiftool.org/}, \texttt{Metagoofil\footnote{https://tools.kali.org/information-gathering/metagoofil}} and \texttt{pdfxplr} to name a few.
%\footnote{https://github.com/sowdust/pdfxplr}
We have analyzed the metadata of PDF files of our dataset using  \texttt{exiftool} and \texttt{pdfxplr}. We assume that the metadata can be 100\% trusted. Some values can be decoys but we cannot detect such situation. Table~\ref{tab:exif} shows the metadata of a PDF file. During our analysis, we noticed that number of metadata fields and their names depend on the PDF producer tool used. Some software also provide an option to the user to either add or remove metadata fields while creating a PDF file.

%%%%%%%%%%%%%%%%%%%%%%%%%%%%%%%%%%%%%%%%%%%%%%%%%%%%%%%%%%%%%%%%%%%%%
\begin{table}[ht]
\renewcommand{\arraystretch}{1.2}
\begin{tabular}{|ll|}
\hline
ExifTool Version Number         &: 11.49\\
File Name                       &: 127-Zadost-2-10-2018.pdf\\
Directory                       &: ./www.vzcr.cz\\
File Size                       &: 259 kB\\
File Modification Date/Time     &: 2019:04:04 13:21:11+02:00\\
File Access Date/Time           &: 2020:08:21 13:22:52+02:00\\
File Inode Change Date/Time     &: 2020:07:31 11:23:04+02:00\\
File Permissions                &: rw-r--r--\\
File Type                       &: PDF\\
File Type Extension             &: pdf\\
MIME Type                       &: application/pdf\\
PDF Version                     &: 1.5\\
Linearized                      &: No\\
Page Count                      &: 1\\
Language                        &: cs-CZ\\
Tagged PDF                      &: Yes\\
Title                           &: MINISTERSTVO OBRANY\\
Author                          &: chocholaty\\
Creator                         &: Microsoft Word 2010\\
Create Date                     &: 2019:04:04 13:16:51+02:00\\
Modify Date                     &: 2019:04:04 13:16:51+02:00\\
Producer                        &: Microsoft Word 2010\\

\hline\hline
\end{tabular}
\caption{Exiftool output for a PDF file in our dataset.}
\label{tab:exif}
\end{table}

\subsection{Positioning of our work}

The different types of hidden data found in PDF files are summarized in Table~\ref{tab:tools}. As explained previously in this section, the problem of hidden data in PDF files is well-known. For instance, Mendelman has demonstrated in his master thesis~\cite{Mendelman2018} that PDF metadata can be used to footprint an organization. He has gathered 1580 PDF files from 3 organizations and was able to collect printer names, internal domain names, Operating Systems, personal information and producer tool information.

\begin{table}[ht]
\begin{center}
\renewcommand{\arraystretch}{1.2}
\begin{tabular}{|r|l|l|}\hline
       \textbf{Hidden data}           & \textbf{Previous Use} & \textbf{Tool} \\ \hline\hline
Dynamic resources & \cite{Castiglione2010}        & N/A \\
Path and Images            & \cite{Aura2006,Garfinkel2014,Feng2018}       &  \texttt{pdfxplr}\\
Annotations           & & \texttt{strings}\\
Comments and object  & & \texttt{grep} \\ 
Metadata          &  \cite{Alonso2009a,Alonso2009b,Aura2006,Garfinkel2014,Mendelman2018}       & \texttt{exiftool}\\\hline
\end{tabular}
\caption{Hidden data and tools needed to recover them.}\label{tab:tools}
\end{center}
\end{table}

Our work can be distinguished from the previous works by the scale of our analysis. We have gathered PDF files produced by the same authors during several years. It allows us to observe authors and even organization's update policies. Moreover, we have been able to observe how security agencies sanitize their PDF files. To the best of our knowledge, this is the first study that measures quantitatively and qualitatively sanitization of PDF files.  
\section{PDF Files Sanitization}\label{sec:sani}

In this section, we describe several PDF sanitization tools that we have tested during our work and then we present four different levels of sanitization methods used by authors. 

\subsection{Sanitization tools}

\textbf{Adobe Acrobat tool}: is often mentioned in NSA guidelines~\cite{NSA2005,NSA2008,NSA2015} as a reliable sanitization tool. It cleans the metadata and all the hidden content of the PDF file. This is the most complete sanitization tool we have used in our work.

\textbf{GhostScript}: Converting PDF file to postscript is mentioned online as one of the sanitization method. This conversion can be done using \texttt{GhostScript} tool (Listing~\ref{listingghost}) and we have observed that it clearly removes a lot of hidden information on the resulting PDF file.   However, it is difficult to determine what is removed or kept by this conversion.  

\begin{center}
\lstset{ %
    basicstyle=\ttfamily\footnotesize,
    frame=single,
    keywordstyle=\color{blue},
    language=Bash,
    showstringspaces=false,
    caption={PDF metadata sanitization using Ghostscript.},
    captionpos=b,
    label=listingghost,
    morekeywords={blue},
}
\begin{lstlisting}[basicstyle=\small]
pdf2ps filename.pdf
ps2pdf filename.ps
\end{lstlisting}
\end{center}

\textbf{Exiftool}: Several threads on sanitization in online forums also mention the possibility to sanitize a PDF file using  \texttt{exiftool} (\url{https://exiftool.org/}). We observed that, it  only cleans the metadata of a PDF file (Listing~\ref{listingex}) and not the rest of the hidden information. 

\begin{center}
\lstset{ %
    basicstyle=\ttfamily\footnotesize,
    frame=single,
    keywordstyle=\color{blue},
    language=Bash,
    showstringspaces=false,
    caption={PDF metadata sanitization using exiftool.},
    captionpos=b,
    label=listingex,
    morekeywords={blue},
}
\begin{lstlisting}[basicstyle=\small]
exiftool -all:all= file.pdf
\end{lstlisting}
\end{center}

\textbf{Text processing software}: can also produce PDF files without any metadata. This is the case for \texttt{Microsoft Office Word} or \texttt{LibreOffice}. \\

\textbf{\LaTeX tools}: It is also possible to include options in \LaTeX sources to remove the metadata (see Listing~\ref{listinghyperref}) or use the \texttt{pdfprivacy} package\footnote{\url{https://ctan.org/pkg/pdfprivacy}}. 

\begin{center}
\lstset{ %
    basicstyle=\ttfamily\footnotesize,
    frame=single,
    keywordstyle=\color{blue},
    language=Bash,
    showstringspaces=false,
    caption={Producing PDF file without metadata with \texttt{hyperref in \LaTeX{} sources}.},
    captionpos=b,
    label=listinghyperref,
    morekeywords={blue},
}
\begin{lstlisting}[basicstyle=\small]
\usepackage{hyperref}
\hypersetup{
pdftitle={},
pdfauthor={},
pdfproducer={},
pdfcreator={},
pdfkeywords={},
}\end{lstlisting}
\end{center}

\subsection{Levels of sanitization}
During our analysis of PDF files of security agencies, we observed that different authors have used different levels of sanitizations. Hence, we distinguish four different levels of PDF file sanitization:

\begin{itemize}
\item \textbf{Level-0} consists of PDF files that include complete metadata information. There is no sanitization. 
\item \textbf{Level-1} consists of PDF files with partial metadata. Some metadata fields have been removed.
\item \textbf{Level-2} consists of PDF files without any metadata. They have been sanitized using  \texttt{exiftool} or by directly producing PDF files without any metadata.
\item \textbf{Level-3} consists of PDF files with no information leakage and properly cleaned. All the objects within the PDF file holding sensitive information have been removed. This level can be obtained using Adobe Acrobat tool.
\end{itemize}

When Level-2 and Level-3 sanitization are observed, we know that the authors of the PDF file have a clear will to sanitize their PDF files. Level-0 and Level-1 are observed when the authors have not applied any sanitization method on their PDF files. 

Solution based on \texttt{exiftool} and by producing PDF files without metadata (Level-2) are not as strong as Adobe Acrobat with respect to sanitization (Level-3). Therefore, as suggested by NSA, Adobe Acrobat tool should be considered as the only reliable solution for complete PDF file sanitization (Level-3).

\section{Exploitation of Hidden Data}\label{sec:exploit}

The National Security Agency (NSA) has high security standards and it has provided many guidelines to sanitize PDF files~\cite{NSA2005,NSA2008,NSA2015}. This was a motivation in our work to check if it was possible to exploit the hidden data in the PDF files published by security agencies around the world. We also wanted to know if anybody follow NSA guidelines on PDF sanitization. 
  
\begin{figure*}[!h]
  \includegraphics[width=17cm, height=12cm]{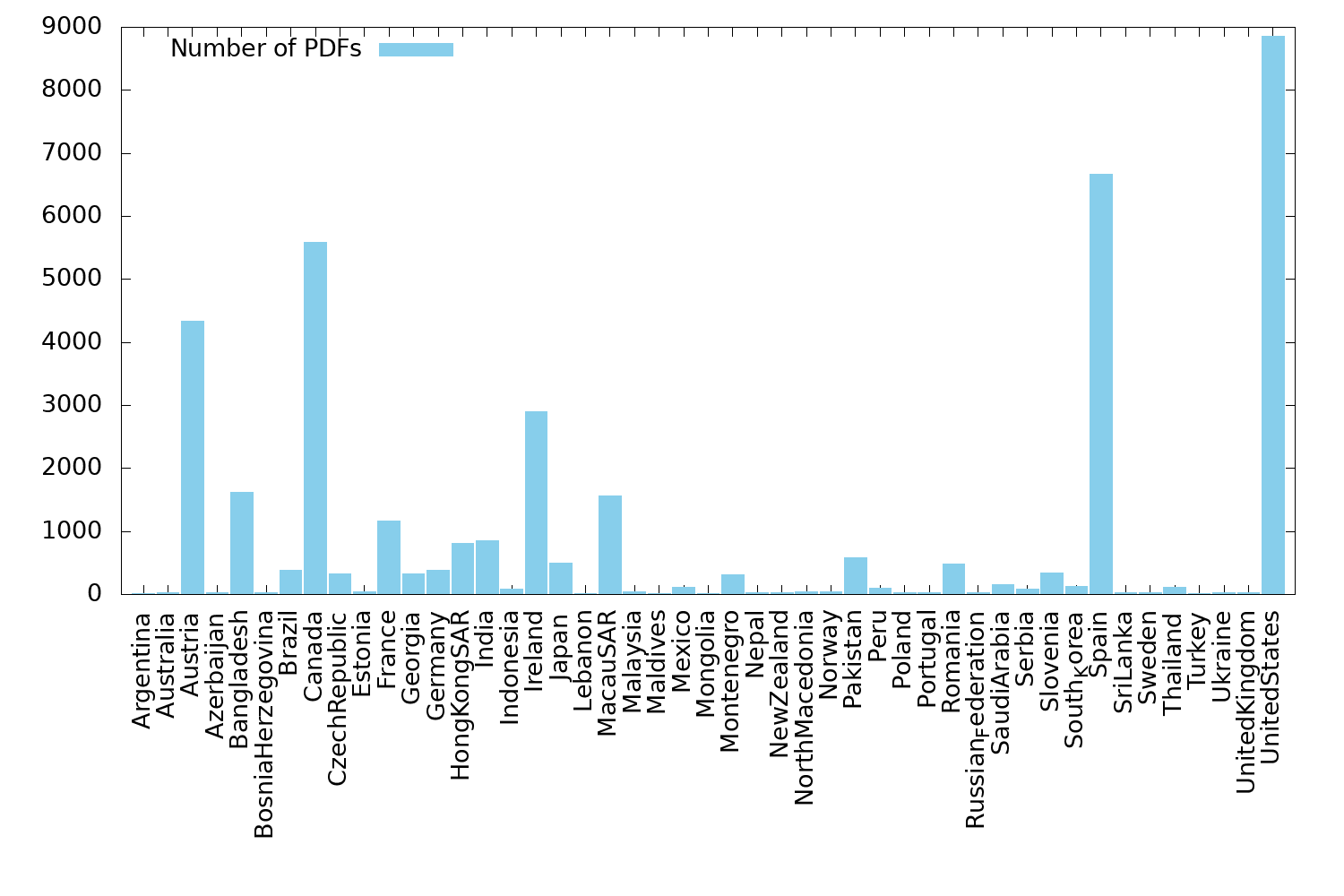}
  \caption{\# PDF files for each country in our dataset.}
  \label{fig:dataset}
\end{figure*}

We have crawled the websites of 75 security agencies mentioned by Wikipedia\footnote{\url{https://en.wikipedia.org/wiki/Security\_agency}} belonging to 47 countries. We have downloaded 39664 PDF files in total. The distribution of PDF files over the agency is not even. We found between 5 to around 6000 PDF files for each agency. Figure~\ref{fig:dataset} shows the discrepancy in the number of PDF files for each country in our dataset. We have used \texttt{wget} command to crawl websites and download these PDF files.
 
We have analyzed the content of the PDF files in our dataset to find if the names of the authors appear directly within the document or not. Using text  extraction commands, we found the author names in 1783 (4\%) PDF files and the rest of the 37881 (96\%) PDF files are anonymous.

It is important to notice that we did not know if the authors of the PDF files were some employees of the security agencies or if they were working for agencies contractors.  We do not know the ground truth, we have made the  assumption that the authors of the PDF files were working for a security agency for simplicity.

In the rest of this section,  we describe different information that we were able to extract on the author and on the organization from PDF files.

\subsection{Information leaked on the authors}
%We report all the information extracted from the PDF files on the author of security agencies.
\textbf{Name of the author}: During our analysis of PDF files, we observed that three metadata fields \texttt{author}, \texttt{creator} and \texttt{Tag Author Email Display Name} reveal the name of the author producing the PDF file. In our dataset, 13166 (33\%) PDF files reveal the identity of the individual who have created the file. \\

\textbf{PDF producer tool used by the author}: Metadata fields like \texttt{producer},  \texttt{creator} and \texttt{creator tool} reveal the name of the PDF producer tool used by the author. We found that, in our dataset 30155 (76\%) PDF files  include the metadata information on the PDF producer tool used. In Table~\ref{Ptool}, we provide the list of popularly used PDF producer tools.  \texttt{Acrobat Distiller}, \texttt{Microsoft Office Word} and \texttt{Adobe PDF Library} are the most popular tools in our dataset.

\begin{table}[H]
\begin{center}
\renewcommand{\arraystretch}{1.2}
\begin{tabular}{|l|l|l|}
\hline
Producer tool & \# PDF& \# Agencies \\
\hline\hline
%Rare producer tools & 18218\\

Acrobat Distiller 		& 9054 (23\%)	&46\\
Adobe PDF Library 		& 6171 (16\%)	& 50\\
Microsoft Office Word   & 4850 (12\%)	& 66\\
LibreOffice 			& 2171	(5\%) & 07\\
Ghostscript 			& 1133 (3\%) & 36\\
Mac OS X Quartz 		& 94 (0.2\%)	& 20\\
SKia/PDF 				& 106 (0.2\%)	& 08 \\
Other tools 			& 16085 (40.5\%) & 75\\\hline
\end{tabular}
\caption{\# of PDF files associated to popular PDF producer tools (76\% PDF files).}
\label{Ptool}
\end{center}
\end{table}

\textbf{Operating System used}: Interestingly, producer tool name in the metadata sometimes reveals the Operating System (OS) used by the author. The reason is that some tools are  OS specific: \textit{Mac OS X 10.6.6 Quartz PDFContext, Acrobat Distiller 20.0 (Macintosh), Acrobat Distiller 8.3.1 (Windows) or Antenna House PDF Output Library 6.2.553 (Linux64)}. In our dataset atleast 16805 (42\%) of the PDF files reveal Operating System information. Table~\ref{tab:osp} shows the distribution of PDF files between the three main Operating Systems: Microsoft Windows, Mac OS and Linux. We can see that Microsoft Windows is a popular choice among the employees of many agencies.

\begin{table}[H]
\begin{center}
\renewcommand{\arraystretch}{1.2}
\begin{tabular}{|l|l|l|}
\hline
OS used & \# agencies & \# PDFs\\ \hline\hline
Microsoft Windows & 71 & 11,174 (28\%)\\
Mac OS&  29 & 3,444 (8\%)\\
Linux & 7 & 2,187 (6\%)\\
\hline
\end{tabular}
\caption{\# of agencies and OS used.}
\label{tab:osp}
\end{center}
\end{table}

\textbf{Brand of the device, e-mail and Path information}: We observed that sometimes authors reveal the brand of their hardware device in the metadata fields author, creator tool and creator instead of their name or along with their name. Using simple string extraction commands, we found four brands like \texttt{Toshiba, HP, DELL and  Lenovo}. Authors of 24 security agencies have such practices (Table~\ref{stats}).

Using the \texttt{pdfxplr tool}, we could also extract different information like the e-mail address of authors, the PATH or location of the folder from where images/files were inserted within the PDF files. We found personal or official e-mail addresses in the PDF metadata. There were 52 unique e-mail addresses in one of the metadata field such as \texttt{Tag Author Email, author} and  \texttt{Current User Email}\ldots 47 of these e-mail addresses are official that belong to the agency where the author works and four of them are \texttt{gmail addresses}, one of them is \texttt{outlook address} (Table~\ref{stats}).

\begin{table}[H]
\begin{center}
\renewcommand{\arraystretch}{1}
\begin{tabular}{|l|l|l|}
\hline
OS & \# PDF & \# Agencies \\
\hline\hline
E-mail & 52 & 13\\
Hardware brand & 581 & 24 \\
Paths & 1814 & 47\\
\hline
\end{tabular}
\caption{\# of PDF files revealing e-mail, hardware and PATH information.}
\label{stats}
\end{center}
\end{table}
%%%%%%%%%%%%%%%%%%%%%%%%%%%%%%%%%%%%%%%%%%%%%%%%%%%%%%%%%%%%%%%%%%%%%
 %PDF files published on the websites of security agency consists of different authors.

Images/files included within the PDF also contain the metadata and the path from where the image was included. Image metadata is out of scope of this paper. We found complete location of where a file is located for 1814 PDF files (Table~\ref{stats}). \\

\textbf{Author's behavior w.r.t software updates}: \emph{It is possible to combine author, producer and time information provided in PDF  metadata fields to understand how employees in security agencies update or change their PDF producer tool}. We have provided three examples in Table~\ref{tab:3author}. Author-X is working at Federal Investigation Agency (Pakistan) and he/she has never changed/updated his/her PDF producer software from 2014 to 2019. This author is using \texttt{Microsoft Office Word 2007} software which is a older version of the software and may contain vulnerabilities that could be exploited. Author-Y works for the Spanish Ministry of Defense and updates his/her \texttt{Acrobat Distiller software} on a regular basis. Author-Z is working at the Customs and Excise Department (Hong Kong) and produced PDF files using Adobe Acrobat. Author-Z sometimes used \texttt{PDFCreator} tool to convert documents into PDF files on \texttt{Microsoft Windows Operating System}. 

%---------------------------------------------------------------------%
%---------------------------------------------------------------------%
\begin{table*}
\begin{center}

\begin{tabular}{|l|l|l|r|l|r|}
\hline
Agency & Author Name & Author habits & Year &PDF producer tool &\# PDF published\\\hline\hline
\textbf{fia.gov.pk} &\textbf{Author-X} &\textbf{Using same tool} & \textbf{2014-19} & \textbf{Microsoft Office Word 2007} &\textbf{ 29}\\\hline
defensa.gob.es&	Author-Y & Updating regularly & 2010&Acrobat Distiller 7.0.5 (Windows) & 4\\
& 	&	& 2011 & Acrobat Distiller 8.0.0 (Windows)&1\\
&&& 2011-14 & Acrobat Distiller 8.2.5 (Windows)&9\\
%&&& 2012 & Acrobat Distiller 8.2.5 (Windows)&2\\
%&&& 2013 & Acrobat Distiller 8.2.5 (Windows)&4\\
%&&& 2014 & Acrobat Distiller 8.2.5 (Windows)&1\\
&&& 2014-15 & Acrobat Distiller 10.1.0 (Windows)&26\\
%&&& 2015 &Acrobat Distiller 10.1.0 (Windows)&9\\
%&&& 2017-18 & Acrobat Distiller 11.0 (Windows)&1\\
&&& 2017-18 & Acrobat Distiller 11.0 (Windows)&3\\\hline
customs.gov.hk& Author-Z & Changing tools & 2017 & Adobe Acrobat 11.0.20&1\\
&&& 2018 & Adobe Acrobat 11.0.0&1\\
&&& 2019 & PDFCreator 2.1.2.0&3\\
&&& 2019 & PDFCreator 3.2.2.13517&2\\
&&& 2019-20 & Adobe Acrobat Standard 2017 17.11.30150&2\\
\hline
\end{tabular}
\caption{Interesting author behaviors observed using PDF producer tool information.}
\label{tab:3author}
\end{center}
\end{table*}

\subsection{Information leakage on the organization}
Now we  focus on the information that could be exploited on the organization level using PDF metadata. During our analysis we observed that, many agencies include more than one author publishing the PDF files. It is possible to download all the PDF files published on a security agency's website and observe the author habits, OS trends\ldots

\begin{figure*}[ht]
\begin{center}
  \includegraphics[width=19cm,height=14cm]{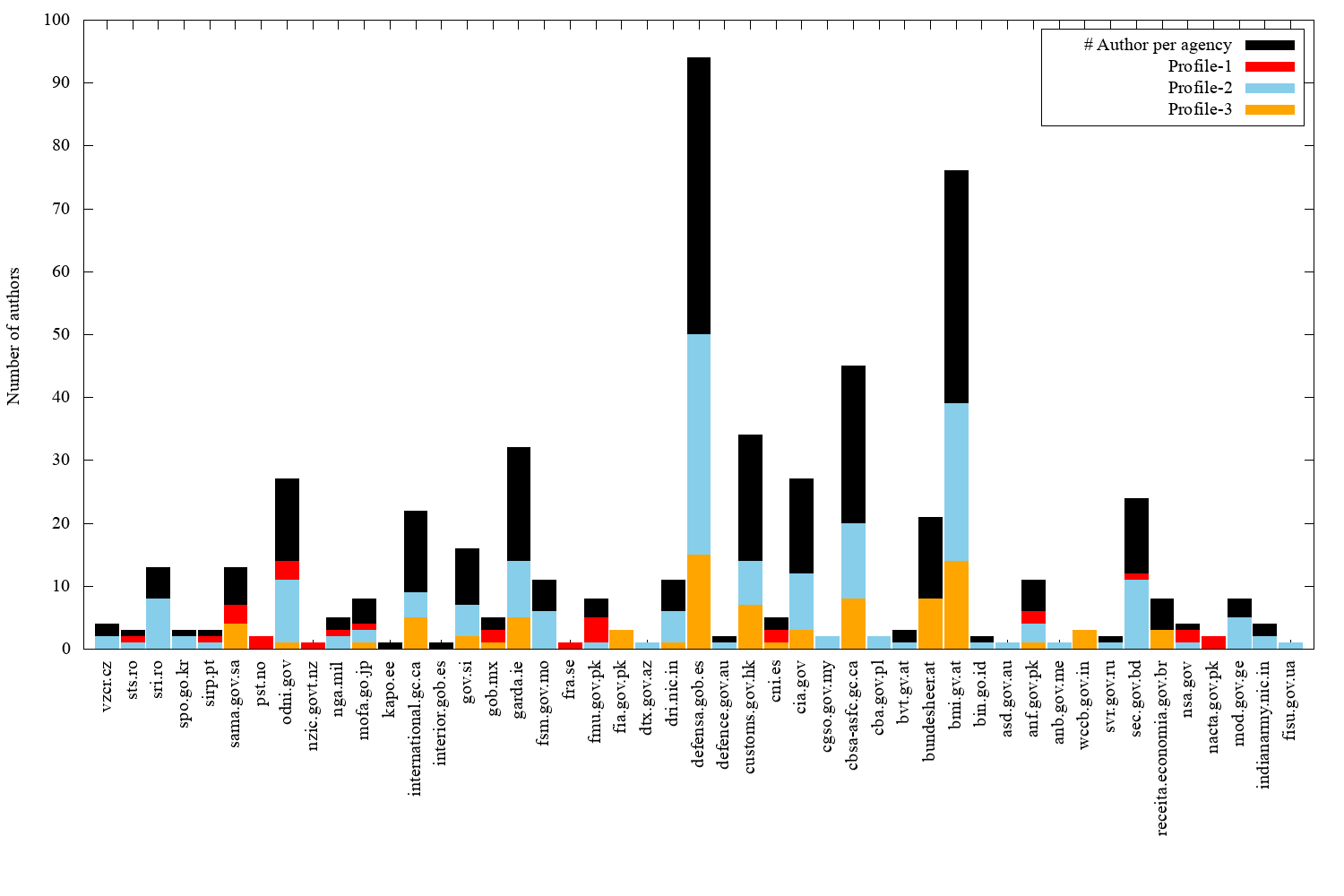}
  \caption{Profile of the employees working in security agencies.}
  \label{fig:AP_stats}
\end{center}
\end{figure*}

By  aggregating the PDF files published for several years, we have considered three different profiles for employees on the organization level (behaviors previously shown in Table~\ref{tab:3author}). \textbf{Profile-1} employees update their software on a regular basis. \textbf{Profile-2} employees have changed their software. Finally, \textbf{Profile-3} employees do not change their software during a period of at least 2 years. Figure~\ref{fig:AP_stats} shows the number of employees  following  \textbf{Profile-1}, \textbf{Profile-2} or \textbf{Profile-3} for each security agency. In our dataset, we found atleast 19 agencies that have 154  employees following \textbf{Profile-3} and who do not change/update their tools over a period of two years or more.\\ %\textcolor{red}{How many employees does it represent?}

\begin{figure}[H]
  \includegraphics[width=9cm,height=7cm]{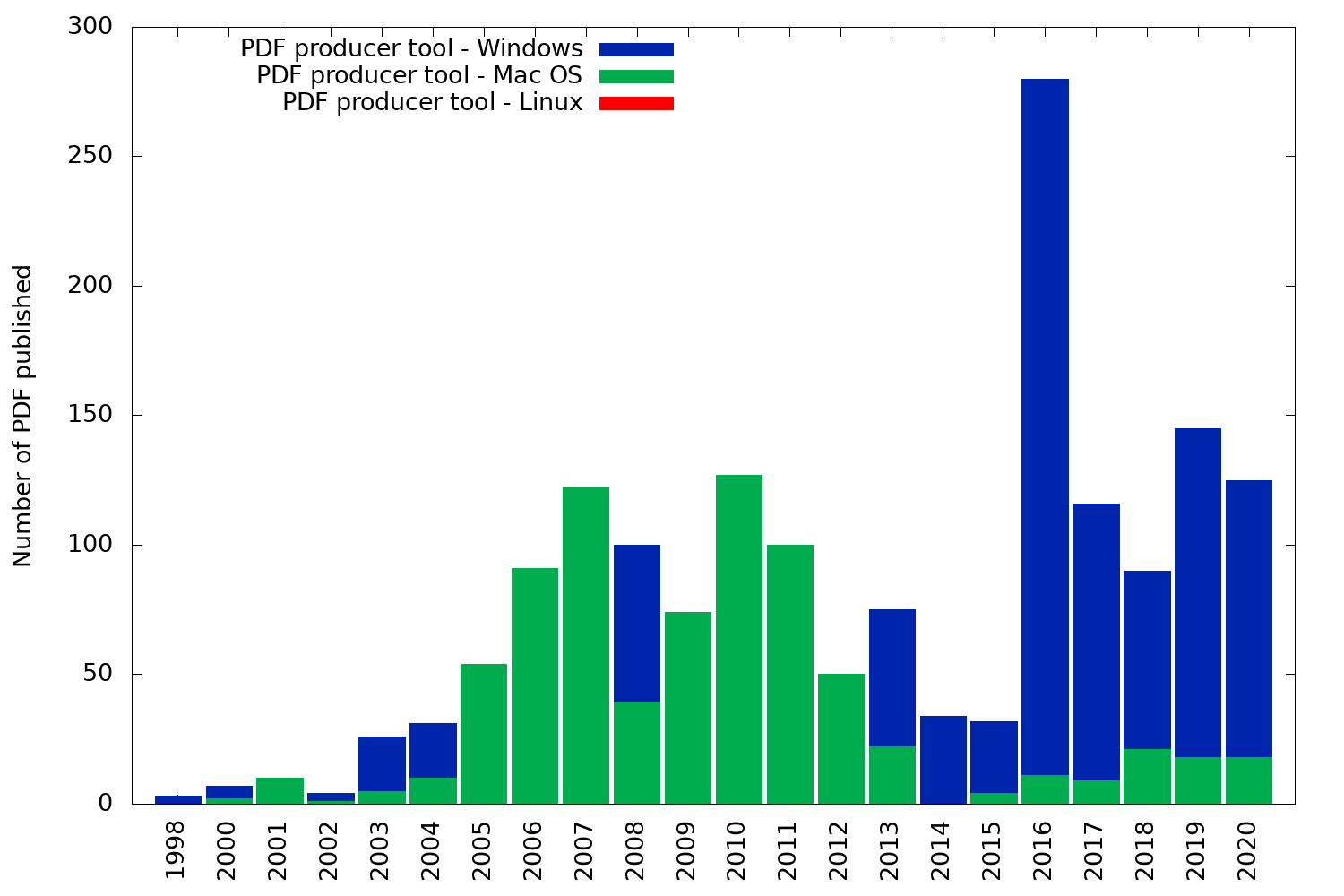}
  \caption{Use of different OS over years at \url{bmi.gv.at}.}
  \label{fig:bmi_os}
\end{figure}

\textbf{OS used by agencies}:  In our dataset, OS details are revealed in 16805 (42\%) PDF files. Table~\ref{tab:osp} shows the distribution of PDF files between the 3 main Operating Systems: Microsoft Windows, Mac OS and Linux and also how many agencies use each of these Operating Systems. It is also possible to spot how many Operating Systems are used in a security agency. We give the example of Austria's Interior Ministry (\url{bmi.gv.at}). Figure~\ref{fig:bmi_os} shows the different Operating Systems we have been able to spot during the last 24 years. In the last five years, the authors of this agency mainly used Microsoft Windows OS to produce their PDF files and fewer PDF files have also been created using MAC OS.

%\textcolor{red}{following paragraph was re-written and Table 7 fixed}\\
\textbf{OS Trends}: Using the OS information, we were able to spot if all the employees working for a agency use same Operating System or there exists a diversity and employees use different OSs.  Table~\ref{tab:os_yearly} shows the use of OSs by agencies for a period of 20 years. Even though the number of agencies using same OS is more, we can observe that in the last five years, number of agencies leaning towards using multiple OSs are increasing. 

%\textcolor{red}{PDF files published on the security agency websites can reveal the information of change of OS. Table~\ref{tab:os_yearly} shows different OS used by agencies for periods of 5 years. In the Table~\ref{tab:os_yearly}, changing OS means use of one OS for a period and then completely switching to another OS. Whereas multiple OS means, randomly using different OSs to produce PDF files. We can observe that many agencies have the habits to use same OS for several years. And in the last five years 25 agencies have started using multiple OSs to publish their PDF files.}

\begin{table}[H]
\begin{center}
\renewcommand{\arraystretch}{1.2}
\begin{tabular}{|l|c|c|}
\hline
year & \# agencies using & \# agencies using \\
& same OS & multiple OS (mix)\\ \hline\hline
2000-2005& 15&9\\
2006-2010&18&10\\
2011-2015&33&17\\
2016-2020&37&26\\
\hline
\end{tabular}
\caption{\# of agencies and OS used.}
\label{tab:os_yearly}
\end{center}
\end{table}

The results obtained from PDF metadata shows that we can learn different information on the author level and also on the organization level by just using the metadata of the PDF files. 
%To avoid information leakage there is a strong need for PDF files to be sanitized. 

%In the next section, we discuss the limit of Level-2 PDF sanitization. All the results presented in this section shows that PDF files include sensitive information and they need to be sanitized.
 
\section{Security Agencies and Sanitization}\label{sec:dataset}

We have evaluated each PDF file in our dataset and extracted its metadata to check the level of sanitization. Table~\ref{sani_level} provides the number of PDF files for each level of sanitization in our dataset. We found that a total of 9509 (24\%) PDF files have been sanitized before being published online. Clearly, PDF sanitization is a concern for several security agencies. However, we found that only 3313 (8\%) PDF files in our dataset were sanitized with Level-3. 

\begin{table}[!ht]
\begin{center}
\renewcommand{\arraystretch}{1.2}
\begin{tabular}{|l|l|l|}
\hline
Level of sanitization & \# PDF \\
\hline\hline
Level-0 sanitization		& 16199 (41\%)\\
Level-1 sanitization		& 13956 (35\%)\\
Level-2 sanitization		& 6196 (16\%)\\
Level-3 sanitization		& 3313 (8\%)\\\hline
\end{tabular}
\caption{Different levels of sanitization used on PDF files.}
\label{sani_level}
\end{center}
\end{table}

We have computed a score for each agency based on the level of sanitization of the PDF files published. This score is the weighted sum of the number of PDF files sanitized with a certain level times the corresponding level. The value $n_i$ is the number of PDF files sanitized with Level-$i$ for $0\leq i \leq 3$ (see Equation~\eqref{eqn:rank}). The highest possible score is 3 and it can only be achieved if the agency publishes all the PDF files with Level-3 sanitization. 

\begin{equation}\label{eqn:rank}
\textrm{Score} = \frac{0 \times n_0 + 1 \times n_1 + 2 \times n_2 + 3 \times n_3}{n},
\end{equation}

For instance, we have downloaded $n=82$ PDF files on \url{nsa.gov} agency. We found that $n_0=45$ (Level-0),  $n_1=24$ (Level-1), $n_2=13$ (Level-2) and $n_3=0$ (Level-3). Therefore, NSA has a score of 0.60.\\
\\
Table~\ref{tab:rank} shows the score distribution of the security agencies.  One security agency (nabis.police.uk) does not care to sanitize any PDF files before publishing. On the other side of the scale, we found no agency with the perfect score. We found only 7 agencies with a score greater or equal to 2: this implies that most of their PDF files are sanitized. Four of these agencies \url{ssi.gouv.fr}, \url{bmi.bund.de}, \url{interior.gob.es} and \url{secp.gov.pk} have performed Level-2 sanitization on most of their PDF files. And atleast, three agencies \url{sie.ro}, \url{garda.ie} and \url{bvt.gv.at} have taken care to sanitize most of their PDF files with Level-3 sanitization. 

\begin{table}[!ht]
\begin{center}
\renewcommand{\arraystretch}{1.2}
\begin{tabular}{|l|l|l|l|l|l|l|l|}
\hline
Score & 0 & 0 \textgreater 1 & 1 & 1 \textgreater 2 &  2 & 2 \textgreater 3 & 3  \\
\hline\hline
\# agencies & 1&50&6&11 &4&3&0\\
\hline
\end{tabular}
\caption{Sanitization score of security agencies.}
\label{tab:rank}
\end{center}
\end{table}

\subsection{Comparative data}

In order to put the results obtained for security agencies into perspective, we have also downloaded more than 500000 PDF files from scientific repositories (Cryptology ePrint Archive\footnote{https://eprint.iacr.org/}  and HAL\footnote{https://hal.archives-ouvertes.fr/} with the agreement of the system administrators of these websites. 

\emph{Are scientists concerned by PDF sanitization?} They are at least when they submit their work to a venue which enforce a double-blinded peer review process. Both reviewers and authors are anonymous from each other throughout the review.   Many conferences (47 in computer science according to \url{http://double-
blind.org}) have enforced a double-blind review submission process. 7 conferences in 2020 (ACM - ASPLOS 2020, HPCA 2020, ICSE 2020, ISCA 2020, MICRO 2019, SIGMOD 2020 and CHI 2020) mentioned explicitly that the authors must remove all identifiable information from the PDF file. ICSE conference also provides instructions to check the metadata using \texttt{pdfinfo} or Adobe Acrobat. But none of these 7 conferences provide instructions on how to clean metadata and sanitize the PDF file before submission. Among the 500000 PDF files downloaded, we found only one PDF file that has been sanitized (level-2). \emph{Clearly, security agencies are more concerned by PDF files sanitization than scientists.}

\subsection{Why level-2 sanitization is not enough?}
As previously mentioned in Section~\ref{sec:sani}, Level-2 sanitization is done either using \texttt{exiftool} or by directly producing the PDF without any metadata information. We observed that, if \texttt{exiftool} is used to sanitize a PDF file, it is still possible to recover all the metadata. 

Metadata are stored in a separate object within a PDF file and \texttt{exiftool} only removes the reference to this metadata object in the file. Hence, it is still possible to access this object. 
Accessing each field of the metadata requires only the use of the \texttt{grep} command. In our dataset, we found 1237 (3\%) PDF files sanitized using \texttt{exiftool} and we were able to recover metadata information using \texttt{grep} command for all these 1237 (3\%) PDF files. \\

We also found that 4959 (13\%) PDF files have been produced without any metadata. This sanitization method is more efficient but we succeed to extract some sensitive information hidden within the PDF. During our analysis of PDF files, we observed that apart from metadata object, some other objects present in PDF can include hidden information. Using the \texttt{grep} command, we performed simple search on the PDF files and found hidden information on the software used by the author producing the PDF file.
% related to the software used by the authors.

Security agencies are recommended to follow proper sanitization methods to avoid leakage of any kind of information within their PDF files.
\section{Conclusion}
Our study shows that the PDF files published by different security agencies are not sanitized to the level expected by such organizations. Many PDF files published by these agencies contained hidden information which can be used to target their employees. We were even able to find an employee who has not updated his/her software during 5 years. Footprinting an organization using its published PDF files is quite effective. 

Some  agencies care about sanitization but only 3 agencies out of 7 are doing it properly. The issue is that popular PDF producer tools are keeping \textbf{metadata by default} with many other information while creating a PDF file. They provide no option for sanitization or it can only be achieved by following a complex procedure.  Software producing PDF files need to enforce \textbf{sanitization by default}. The user should be able to add metadata only as an option.

In particular, our study urges for security agencies to take measures that should enforce stronger sanitization methods to limit the risk of information leak in their PDF files.

\section{Acknowledgments}
This work has been supported by the SIDES 3.0 project (ANR-16-DUNE-0002) funded by the French Program Investissement d'Avenir and the Grenoble Alpes Cybersecurity Institute CYBER@ALPS under contract ANR-15-IDEX-02. 
\bibliographystyle{ACM-Reference-Format}
\bibliography{ref}

\end{document}